\documentclass[twocolumn,showpacs,amsmath,amssymb,aps,pre]{revtex4-1}

\usepackage{graphicx}
\usepackage{dcolumn}
\usepackage{bm}
\usepackage{xcolor}

\begin{document}

\title{When is a quantum heat engine quantum?}

\author{Alexander Friedenberger}
\affiliation{Department of Physics, Friedrich-Alexander-Universit\"at Erlangen-N\"urnberg, D-91058 Erlangen, Germany}

\author{Eric Lutz}
\affiliation{Department of Physics, Friedrich-Alexander-Universit\"at Erlangen-N\"urnberg, D-91058 Erlangen, Germany}

\begin{abstract}
Quantum thermodynamics studies  quantum effects in thermal machines. But when is a heat engine, which cyclically interacts with external reservoirs that unavoidably destroy its quantum coherence, really quantum? We here use the Leggett-Garg inequality to assess the nonclassical properties of a single two-level  Otto engine. We provide the complete phase diagram characterizing the quantumness of the engine as a function of its parameters and identify three distinct phases. We further derive an explicit expression for the transition temperature.
\end{abstract}


\maketitle
The study  of thermal machines has been a cornerstone of thermodynamics since its origins. The performance of these devices  is usually characterized by two key parameters:  efficiency
and  power \cite{cen01}. While attention has for a long time solely focused on idealized infinite-time
processes---with maximum efficiency but zero power---research in the last decades has
shifted towards more realistic finite-time cycles---with reduced efficiency and non-vanishing power \cite{and84,and11}. Macroscopic internal combustion engines, refrigerators and heat pumps are notable examples of classical thermal machines whose finite-time dynamics follow the laws of classical physics  \cite{cen01,and84,and11}. Quantum heat engines, on the other hand, are microscopic motors that are described by dynamical equations of motion that obey the laws of quantum mechanics \cite{kos14}. They have been extensively studied  over the last fifty years \cite{sco59,ali79,kos84,gev92,gev92a,gor91,fel96,fel00,scu02,kie04,lin03,rez06}, and both quantum coherence \cite{scu03} and quantum correlations \cite{dil09,aba14} have been theoretically shown to be able to boost their performance. Meanwhile, concrete proposals  to experimentally build such quantum machines have  been put forward using, for instance, trapped ions \cite{aba12,ros14} and nanomechanical systems \cite{zha14,dec15}. However, contrary to most quantum  applications that aim at perfectly shielding systems from their environments \cite{nie00,bol91,mau12}, thermal machines, by their very nature, cyclically interact with heat reservoirs that unavoidably destroy their quantum coherence. A crucial issue that therefore  needs to be addressed, for both theoretical and practical reasons, is  how to  identify the quantumness of a  heat engine.

Characterizing nonclassicality   is of fundamental importance in many areas of quantum physics, from quantum computation \cite{zag14,ron14} to quantum biology \cite{lam13,moh14}. Assessing the nonclassical properties of a given state or dynamics is a nontrivial task, however, and, more than a century  after the birth  of quantum theory, the border between classical and quantum worlds remains fuzzy \cite{zur91,leg02,sch07}. A common approach to identify quantum behavior is to impose classical constraints that are violated by quantum mechanics \cite{li12}. The assumptions of realism and locality thus lead to Bell's inequality \cite{bel64}, while those of    macroscopic realism and non-invasive measurements to the Leggett-Garg inequality (LGI) \cite{leg85}. These  two results allow to test whether a system has stronger-than-classical spatial or temporal correlations, respectively. The advantage of the LGI is that it applies to a single particle and not to a pair of particles as required by Bell's inequality. It has therefore become a valuable theoretical and experimental tool to detect quantumness of individual systems \cite{ema14}. In the last few years, experimental violations of the LGI have been observed in an increasing number of single systems, including a superconducting qubit \cite{pal10}, a  spin in a diamond defect  center \cite{wal11},  a nuclear spin \cite{ath11,sou11}, a photon  \cite{gog11,xu11}, a  phosphorus impurity in silicon \cite{kne12}, and a single diffusing atom \cite{rob15}. It is worth mentioning that the last two experiments have implemented ideal  negative measurements, as originally discussed by Leggett and Garg. In the following, we employ the LGI to assess the quantumness of a paradigmatic quantum thermal machine: a single two-level Otto  engine \cite{gev92,gev92a,gor91,fel96,fel00}. Our investigations reveal the existence of three different regimes which we summarize in a phase diagram as a function of the parameters of the engine (see Figs. 3 and 4 below). We further derive an explicit formula for the transition temperature. We finally show  that trying to operate a thermal machine faster, with the idea of keeping  it coherent, actually leads to incoherent dynamics, owing to  constraints imposed by thermodynamics.

\textit{Quantum Otto engine.} We consider a quantum Otto engine for a single two-level system with time-dependent frequency $\omega_t$ and Hamiltonian  $H= \omega_t \sigma_z/2$, where $\sigma_z$ is the usual Pauli operator (we  set $\hbar=k_B=1$). The quantum Otto motor is a generalization of the familiar four-stroke car engine and  its thermodynamic cycle consists of the following isentropic and isochoric steps, as shown in Fig.~1: (1)  \textit{Heating}:  the two-level system is weakly coupled to a hot reservoir at temperature $T_h$ during time $\tau_h$, while its frequency is kept fixed, (2)  \textit{Expansion:} the system is isolated and its frequency is unitarily changed from $\omega_2$ to $\omega_1$ during time $\tau_1$, (3) \textit{Cooling}: the system weakly interacts with a cold reservoir at temperature $T_c$ during time $\tau_c$, while its frequency is again held constant, (4) \textit{Compression}: the frequency is unitarily brought back to its initial value $\omega_2$ during time $\tau_2$.

Work and heat along the different steps may be determined from infinitesimal variations of the average  energy of the system, $E=\langle H\rangle= \omega P$, where  $P= \langle \sigma_z\rangle/2$ is the polarization. We write accordingly $dE= Pd\omega +\omega dP  = \delta W + \delta Q$ and associate work $\delta W$ with changes of frequency and heat $\delta Q$ with changes of polarization. The Otto engine exchanges energy in the form of work during compression/expansion phases and in the form of heat during heating and cooling. In order to compute these quantities along the four branches of the cycle, we specify the finite-time dynamics of an arbitrary  system observable $X_t$  with the Markovian quantum master equation in the Heisenberg picture  \cite{bre07,ali07}, 
\begin{equation}
\begin{aligned}
\frac{dX}{dt}&=i\frac{\omega}{2} \left[\sigma_z,X\right]+\frac{\partial X}{\partial t} \\
&+{\frac{\gamma_{0}}{2}n(\omega,T)}\left(\sigma_{-}\left[X,\sigma_{+}\right] +\left[\sigma_{-},X\right]\sigma_{+}\right)\\
&+{\frac{\gamma_{0}}{2}(n(\omega,T)+1)}\left(\sigma_{+}\left[X,\sigma_{-}\right] +\left[\sigma_{+},X\right]\sigma_{-}\right),\label{eq:mastereq}
\end{aligned}
\end{equation}
where $\sigma_\pm$ denote the raising and lowering spin operators and $n(\omega,T)= 1/[\exp(\beta \omega)-1]$ the bosonic thermal occupation number at inverse temperature $\beta=1/T$. The coupling constant $\gamma_0$ vanishes during the unitary compression/expansion steps  2) and 4) when the engine is not interacting with the heat reservoirs; in this limit Eq.~\eqref{eq:mastereq} reduces to the standard Heisenberg equation.

\begin{figure}[t]
\includegraphics[width=0.39\textwidth]{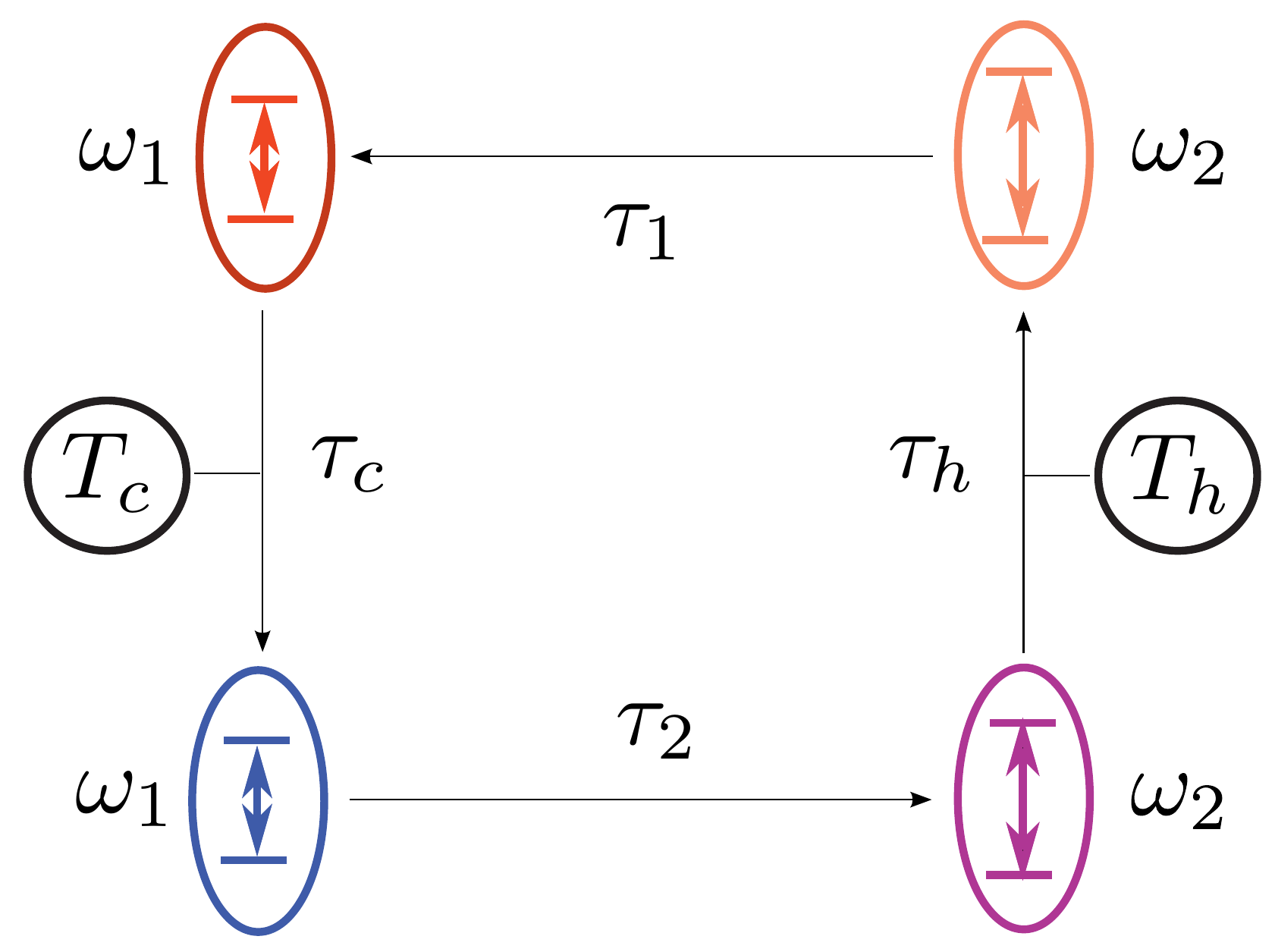}
\caption{(color online) Four branches of the quantum Otto cycle for a two-level system: (1) constant frequency  heating during $\tau_h$, (2) unitary expansion during  $\tau_1$, (3) constant frequency cooling during $\tau_c$, (4) unitary compression during $\tau_2$.   }
\label{fig:LGI}
\end{figure}

During the finite-time operation of the engine, irreversible work is produced along the compression and expansion steps owing to nonadiabatic transitions. We model  the corresponding increase of the polarization with the help of the formula,   $P_t=P_0 +(\sigma/\tau_i)^2 t, (i=1,2)$ \cite{gor91,fel96,fel00}. The phenomenological coefficient $\sigma$ is often referred to as internal friction, as its presence reduces the performance of the engine \cite{gor91,fel96,fel00}. It is directly related to the irreversible entropy production in the usual manner, as shown in \cite{sm}.  The quantum Otto cycle is thus characterized by the following eight parameters, $\omega_{1,2}$, $\tau_{1,2}$, $T_{h,c}$, $\tau_{h,c}$, $\gamma_0$ and $\sigma$. However, not all choices of parameter values are physically meaningful as two important conditions need to be fulfilled in order to have a proper heat engine: the total work produced by the machine  should be positive and the  cycle closed. Thermodynamics therefore restricts the  allowed range of the above parameters, in particular, that of the heating and cooling times $\tau_{h,c}$ and of the internal friction coefficient $\sigma$.

The optimal heating and cooling times $\tau_{h,c}$ were determined in Ref.~\cite{fel00} using the method of Lagrange multipliers. Computing  work and heat  along the four branches of the Otto cycle, assuming, for concreteness,  a linear driving protocol, $\omega_t = \dot \omega t + \omega_0$, the total work produced by the engine in its steady state is found to be \cite{fel00} (the derivation is summarized in \cite{sm} for completeness),
\begin{eqnarray}
\label{2}
W&=&-(\omega_2-\omega_1)\left[ \Delta P^{eq}{\frac{(1-x)(1-y)}{(1-xy)}}\right.\\ 
&-&\left. \frac{\sigma^2}{(1-xy)}\left(\frac{x}{\tau_1}+\frac{1}{\tau_2}\right)(1-y)\right]+\sigma^2\omega_1\left(\frac{1}{\tau_2}+\frac{1}{\tau_1}\right), \nonumber
\end{eqnarray}
where $\Delta P^{eq}= P_h^{eq}-P_c^{eq}=-\tanh(\omega_2/(2T_h))/2+\tanh(\omega_1/(2T_c))/2$ is the difference between the equilibrium polarizations after complete thermalization with the hot and cold reservoirs, $x=\exp({-\gamma_c \tau_c})$ and  $y=\exp({-\gamma_h \tau_h})$ with $\gamma_{h,c} = \gamma_0 \coth(\omega/(2T_{h,c}))$. Optimizing the work \eqref{2} with the constraint of a fixed total cycle time $\tau = \tau_1+\tau_2 +\tau_h+\tau_c$ leads to the Lagrange function,
\begin{equation}
\mathcal{L}(x,y,\lambda)=W +\lambda\left(\tau+\frac{1}{\gamma_c}\ln x +\frac{1}{\gamma_h}\ln y-\tau_1-\tau_2 \right),
\end{equation}
with Lagrange multiplier $\lambda$. Equating the partial derivatives of $\mathcal{L}(x,y,\lambda)$ with respect to $x$ and $y$ to zero  and setting for simplicity $\gamma_h=\gamma_c=\gamma$ eventually yields (keeping the frequencies $\omega_{1,2}$ and times $\tau_{1,2}$ constant) the  optimal hot and cold thermalization times \cite{fel00},
\begin{eqnarray}
\tau_h&=&-\frac{1}{\gamma}\ln\frac{x_\text{max}-\sqrt{Rx_\text{max}(1+R-x_\text{max})}}{x_\text{max}(R+1)},\\
\tau_c&=&-\frac{1}{\gamma}\ln\frac{x_\text{max}-\sqrt{Rx_\text{max}(1+R-x_\text{max})}}{(R+1)},\label{eq:tauh}
\end{eqnarray}
for an engine with vanishing positive work---this is the minimal requirement for a thermal machine to be a heat engine. In that case, the total cycle duration $\tau$ is the minimum time needed to obtain a nonnegative work output. This minimum time vanishes in the limit $\sigma$ to zero \cite{fel00}. The two quantities $R$ and $x_\text{max}$ are defined as,
\begin{equation}
R=\frac{\sigma^2\omega_1\left({1}/{\tau_1}+{1}/{\tau_2}\right)}{\Delta \omega(\Delta P^{eq}+{\sigma^2}/{\tau_1})},\quad
x_\text{max}=\frac{\Delta P^{eq}-{\sigma^2}/{\tau_2}}{\Delta P^{eq}+{\sigma^2}/{\tau_1}}.
\end{equation}
The constraint of a closed cycle, that is, of a finite cycle duration $\tau$, leads to an additional condition on the internal friction coefficient $\sigma$ that reads  \cite{fel00}, 
\begin{equation}
\frac{\sigma^2}{\tau_2}\le \Delta P^{eq} .\label{eq:Bound}
\end{equation}
Heating and cooling times $\tau_{h,c}$ diverge when the above bound is saturated, indicating that large nonadiabatic effects, generated by the finite-time driving, result in increased, and eventually infinite, thermalization times.

Expressions (4) and (7)  form the basis of our investigation of the quantum nature of the Otto engine. Since coherence is preserved along the unitary compression and expansion steps, our strategy is to look for violations of the LGI during the heating phase, where decoherence is expected to be the most adverse. Our analysis may be easily extended to  the cooling phase  as well. 

\textit{Leggett-Garg inequality.} Under the classical assumptions of  macroscopic realism and noninvasive measurability, the two-time correlation functions $C_{ij}$ of a dichotomic observable, $Q=\pm 1$, measured at three distinct times $t_i$, $i=(1,2,3)$,  satisfy the inequality \cite{leg85},
\begin{equation}
K_3=C_{21}+C_{32}-C_{31}\le 1.\label{eq:LGI}
\end{equation} 
Quantum mechanics violates the above inequality. A value of the Leggett-Garg function $K_3$ above one is therefore a clear  signature of nonclassical behavior \cite{ema14}.

In order to assess the quantumness of the Otto engine, we evaluate the symmetrized two-time correlation functions $C_{ij}$ of the observable $\sigma_x$ for the dynamics given by the master equation \eqref{eq:mastereq} with the help of  the quantum regression theorem \cite{bre07}. We find \cite{sm},
\begin{equation}
\!C_{ij}=\frac{1}{2}\langle \left\{\sigma_x(t_i),\sigma_x(t_j)\right\}\rangle=e^{-\frac{\gamma}{2}(t_i-t_j)}\!\cos\left[\omega_2 (t_i-t_j)\right].
\end{equation}
By further considering equally spaced measurement times with time separation $t$, as commonly done \cite{ema14}, we obtain the Leggett-Garg function,

\begin{equation}
K_3(t)=2e^{-\frac{\gamma}{2}t}\cos\left( \omega_2 t\right)-e^{-\gamma t} \cos\left( 2\omega_2 t\right). \label{eq:LGIdamped}
\end{equation}
\begin{figure}[t]
\includegraphics[width=0.48\textwidth]{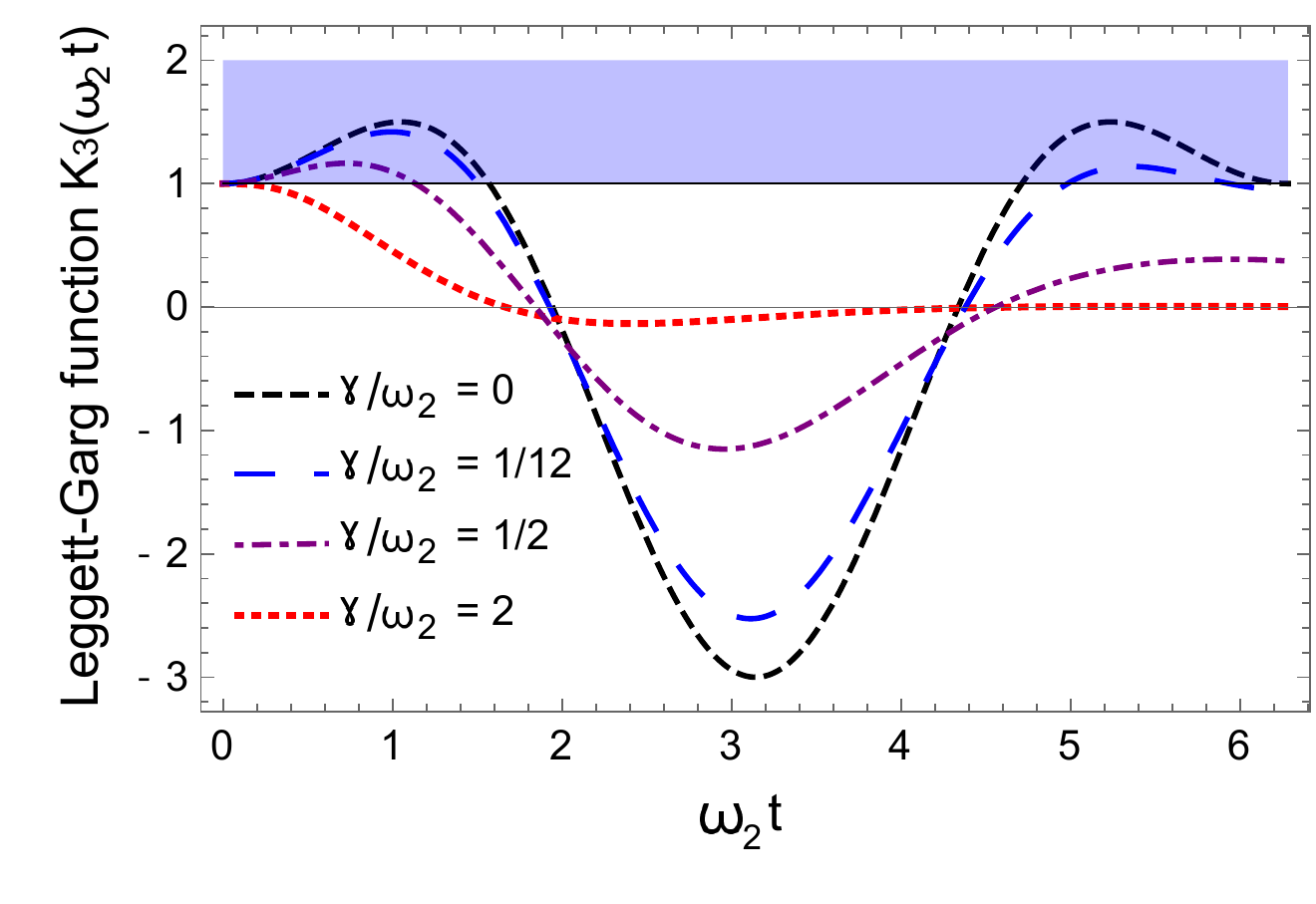}
\caption{(color online) Leggett-Garg function  $K_3(t)$, Eq.~\eqref{eq:LGIdamped}, a function of time for four different values of the damping constant $\gamma$. The shaded area indicates the classically forbidden region $K_3>1$. Above the critical value, $\gamma/2\simeq \omega_2$,  there are no violations of the LGI \eqref{eq:LGI} for any value of $t$.}
\label{fig:LGI}
\end{figure}
The function $K_3(t)$ is shown in Fig. 2 for different values of the damping parameter $\gamma$. We observe that the LGI \eqref{eq:LGI} is violated   for unitary time evolution with $\gamma=0$. Moreover, the amplitude of the violations decreases with increasing $\gamma$ until the critical value, $\gamma/2 \simeq \omega_2$, at which the time scale of the coherent system dynamics  coincides with that of the reservoir induced decoherence,  is reached. Beyond that point there is no violation of the LGI for any value of $t$ and the incoherent evolution imposed by the hot reservoir prevails. In the sequel, we set the first measurement at the beginning of the heating phase and the third measurement at the end, that is, we choose $2t = \tau_h$.

\begin{figure}
\vspace{-.4cm}
\includegraphics[width=.49\textwidth]{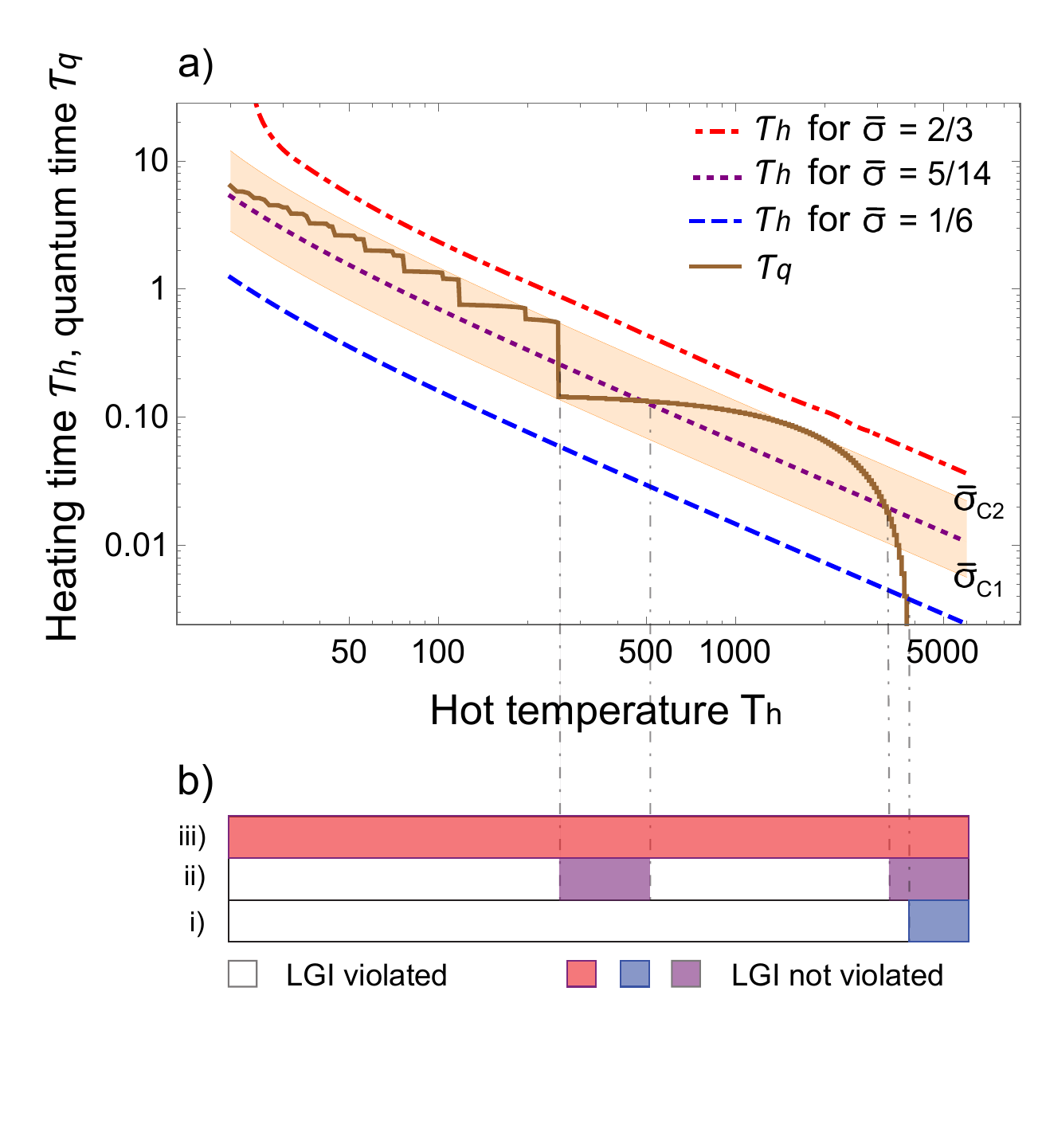}
\vspace{-1.2cm}
\caption{(color online) Phase diagram characterizing  the quantumness of the Otto engine as a function of the hot temperature $T_h$ and the internal friction coefficient $\bar \sigma=\sigma/\tau_2$. a) The engine behaves nonclassically when the LGI \eqref{eq:LGI} is violated, i.e., when  the quantum time $\tau_q$ is larger than the heating time $\tau_h$. b) We identify three regimes: i) single quantum-classical transition for small $\bar \sigma$, ii) multiple transitions in an intermediate domain, iii) classical dynamics  for large $\bar \sigma$. Parameters are $\omega_1 =10$, $\omega_2=20$, $\tau_1 =0.01 $, $\tau_2=0.1$ and $T_c=1$.}
\label{fig:Quantumness}
\end{figure}

We characterize the transition from coherent to incoherent dynamics by introducing  a time $\tau_q$ defined as the largest possible time for which the LGI can be violated:
\begin{equation}
\tau_q=\max\left\{2t|K_3(t)>1\right\}.
\end{equation} 
The factor 2 comes from the fact that   the three measurements span a total time interval of $2t$. Owing to the oscillating feature of the Leggett-Garg function \eqref{eq:LGIdamped}, the quantum time $\tau_q$ is a step function; it decreases with increasing reservoir temperature, as expected, see Fig. 3.

\textit{Results and discussion.} The behavior of the engine depends  decisively on the relationship between heating time $\tau_h$ and quantum time $\tau_q$, that is, on the one hand, on the internal dynamics of the machine, as dictated by thermodynamics, and, on the other hand, on the decoherence process induced by the coupling to the reservoir. The dynamics of the Otto motor is nonclassical only if the quantum time is larger than the heating time, $\tau_q > \tau_h$, so that the LGI may be violated. As seen from Eqs.~(4) and (6), the heating time $\tau_h$ depends on the two temperatures $T_h$ and $T_c$ and on the internal friction coefficient $\sigma$; we will use in the following the reduced quantity $\bar \sigma= \sigma/\tau_2$.

\begin{figure}[t]
\includegraphics[width=.48\textwidth]{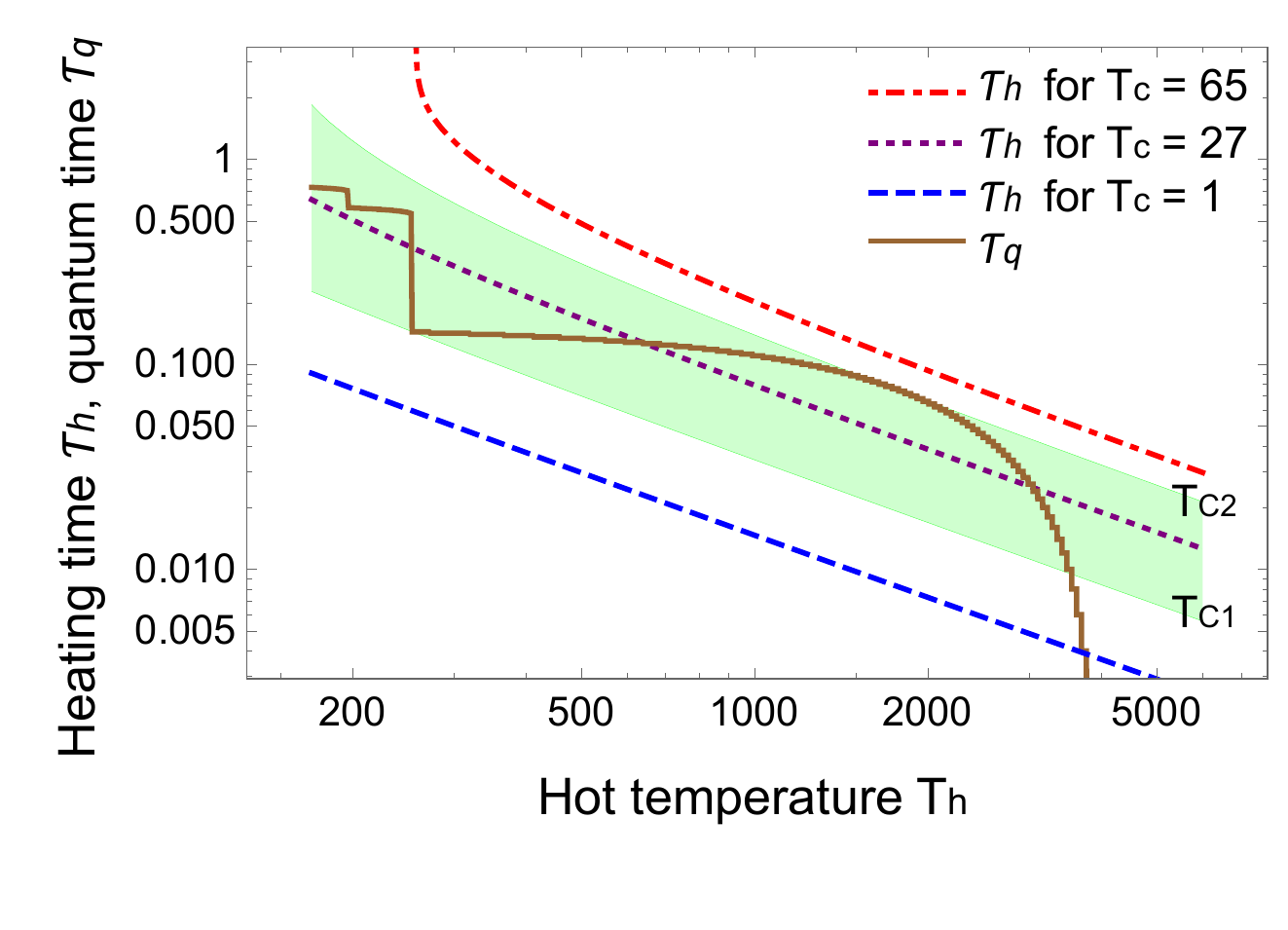}
\vspace{-0.2cm}
\caption{(color online) Phase diagram characterizing  the quantumness of the Otto engine as a function of the hot and cold temperatures $T_h$ and $T_c$. The engine displays an analogous behavior as in Fig. 3, the cold temperature $T_c$ having a similar effect as the internal friction coefficient $\sigma$.  Same parameters as in Fig.~3 with $\bar \sigma = 2/3$.} \label{fig:Quantumness}
\end{figure}

Our main results are summarized in the phase diagram shown in Fig.~3. For a fixed cold temperature $T_c$, we identify three distinct regimes: i) $\bar \sigma < \bar \sigma_{C1}$: for $\bar \sigma$ smaller than a  critical value $\bar \sigma_{C1}$, the behavior of the engine is quantum for temperatures $T_h< T_q= \omega_2/(2\coth^{-1}(2\omega_2/\gamma_0))$; owing to the almost vertical shape of  $\tau_q$, the threshold temperature $T_q$ is mostly independent of $\bar \sigma$ in that domain; ii)    $\bar \sigma_{C1} <\bar \sigma < \bar \sigma_{C2}$: for intermediate values of $\bar \sigma$, the engine appears quantum in some temperature intervals and classical in some others. This corresponds to a transition regime where the boundary between classical and quantum worlds is blurry; iii)     $\bar \sigma_{C2} <\bar \sigma$: for $\bar \sigma$ larger than a second critical value $ \bar \sigma_{C2}$, the dynamics of the Otto engine is classical for all temperatures $T_h$. Note that the tendency of $\tau_h$ to diverge for larger values of $\bar \sigma$ is clearly visible in the upper left corner of Fig.~3a).

The phase diagram in Fig.~3 reveals some subtle interplay between thermodynamics and quantum physics. We first remark  that in the low $\bar \sigma$ phase, the threshold temperature $T_q$ is equivalent to the condition $\gamma/2\simeq \omega_2$ for a two-level system coupled to a single reservoir at temperature $T_2$. Here the internal dynamics of the engine which is set by thermodynamics  does not play any role, and the quantumness  of the Otto cycle is solely controlled by the decoherence process generated by the reservoir. The situation changes dramatically in the two other phases  as the heating time $\tau_h$ increases significantly with increasing $\bar \sigma$. In the high $\bar \sigma$ regime, thermodynamics thus imposes  a purely classical behavior, although decoherence alone would predict the existence of a quantum sector for low enough temperatures.   In other words, the classical or quantum nature of the engine    is not just  governed by the coupling to the reservoirs.  A second, nonintuitive, observation is that trying to operate the engine faster in order to keep it coherent is bound to fail: shortening the compression/expansion steps indeed results in increased nonadiabatic excitations induced by a finite $\bar \sigma$ \cite{rem}, and, in turn, to longer thermalization times. The latter  eventually lead to stronger decoherence and, consequently, to incoherent dynamics. The rule 'the faster, the better' which might be true for most quantum applications, hence does not apply to quantum heat engines.

Figure 4 finally presents the influence of the cold reservoir temperature $T_c$ (for fixed $\bar \sigma$) on the quantumness of the Otto engine. We note that it is largely similar to that of the internal friction $\bar\sigma$ displayed in Fig.~3, as both parameters have an equivalent role in exciting the  system. For $T_c$ below the critical value $T_{C1}$, it does not affect the behavior of the engine which is controlled by the same threshold temperature $T_q$ given above (the discussion in the previous section was based in this $T_c$ regime). In an intermediate  domain, the dynamics of the engine alternates between quantum and classical as $T_h$ is increased.   Ultimately, for $T_c$ larger that the critical value $T_{C2}$,  the LGI is never violated and the engine is purely classical. 

\textit{Conclusions.} We have used the Leggett-Garg inequality to completely characterize the quantum properties  of a two-level  Otto  engine. We have obtained a phase diagram that allows to identify the parameter regimes where the engine behaves nonclassically, and where, therefore, quantum resources in the form of  coherence or   correlations, might be successfully exploited. We have further shown that trying to run a thermal machine faster, with the hope of beating decoherence, is not a winning strategy, as  thermodynamics will make the dynamics classical.

\textit{Acknowledgments.} This work was partially supported by the EU Collaborative Project TherMiQ (Grant Agreement 618074) and the COST Action MP1209.

\section*{Appendix}

\subsection{Relation to entropy production}
We here establish a connection between the internal friction coefficient $\sigma$ and the irreversible entropy production $\Sigma$.
We begin by evaluating the heat exchanged during the two isochoric  steps. Since the frequency is constant, we have in any time interval $[t_1,t_2]$,
\begin{equation}
Q=\int_{t_1}^{t_2}\omega_i \dot{P}_i \:dt =\omega (P_{t_2}-P_{t_1}).
\end{equation}
As a result, we find for the heating and cooling steps,
\begin{eqnarray}
Q_h&=&\omega_2 (P_B-P_A)\\
Q_c&=&\omega_1(P_D-P_C)
=-\omega_1\left(P_B-P_A+\sigma^2\left[\frac{1}{\tau_2}+\frac{1}{\tau_1}\right)\right], \nonumber
\end{eqnarray}
where $A, B, C, D$ denote the four corners of the thermodynamic cycle \cite{fel00}.
The entropy variation of the engine vanishes for a full cycle. The overall entropy change is then given by the entropy difference in the two reservoirs:
\begin{eqnarray}
\Delta S&=&-\left(\frac{Q_c}{T_c}+\frac{Q_h}{T_h}\right)\\
&=&\left(\frac{\omega_1}{T_c}-\frac{\omega_2}{T_h}\right)\left(P_B-P_A\right)+ \frac{\omega_1\sigma^2}{T_c}\left(\frac{1}{\tau_2}+\frac{1}{\tau_1}\right). \nonumber
\end{eqnarray}
The difference in polarization $P_B-P_A$ is further \cite{fel00},
\begin{equation}
P_B-P_A=\Delta P^{eq}\frac{(1-x)(1-y)}{1-xy}-\frac{\sigma^2(1-y)(\frac{x}{\tau_1}+\frac{1}{\tau_2})}{1-xy}
\end{equation}
and we can thus cast the  entropy variation in the  form,
\begin{equation}
\begin{aligned}
\Delta S&=\left(\frac{\omega_1}{T_c}-\frac{\omega_2}{T_h}\right)\Delta P^{eq} \frac{(1-x)(1-y)}{1-xy}\\
&- \left(\frac{\omega_1}{T_c}-\frac{\omega_2}{T_h}\right)\frac{\sigma^2(1-y)(\frac{x}{\tau_1}+\frac{1}{\tau_2})}{1-xy}\\
&+ \frac{\omega_1\sigma^2}{T_c}\left(\frac{1}{\tau_2}+\frac{1}{\tau_1}\right).\label{eq:entropy}
\end{aligned}
\end{equation}
We here identify three sources of irreversibility: partial thermalization ($x,y$ vanish after complete equilibration), the internal friction $\sigma$, and  the intrinsic irreversibility of the  engine given by the term $({\omega_2}/{T_h}-{\omega_1}/{T_c})$. The efficiency of the Otto engine is indeed only equal to the  Carnot efficiency, when ${\omega_1}/{\omega_2}={T_c}/{T_h}$.
In that quasistatic limit, the first two terms in Eq.~\eqref{eq:entropy} vanish and the total entropy change simplify to,
\begin{equation}
\Delta S=\frac{\omega_1\sigma^2}{T_c}\left(\frac{1}{\tau_2}+\frac{1}{\tau_1}\right)= \frac{\omega_1\sigma^2}{T_c\tau_2}+\frac{\omega_2\sigma^2}{T_h\tau_1}. \label{eq:entropyRev}
\end{equation}
We recognize the usual expression for the entropy production in the long-time limit, $\Delta S = \sum_i\Sigma_i/\tau_i$ \cite{esp00} with 
$\Sigma_c={\omega_1\sigma^2}/{T_c}$ and $\Sigma_h={\omega_2\sigma^2}/{T_h}$. A finite friction  coefficient $\sigma$ therefore leads to a finite entropy production.

\subsection{Total work produced by the engine}
For completeness, we summarize in this section the derivation of the  work produced by the engine \cite{fel00}. 
The adiabatic work during either compression/expansion is,
\begin{equation}
W_i=\int_0^{\tau_i}P\dot{\omega}dt=\left(\omega_f-\omega_i\right)\left[\frac{\sigma^2}{2\tau_i}+P_0\right],
\end{equation}
where $\omega_{i,f}$ denotes the initial and final frequencies $\omega_{1,2}$.
Additionally, the irreversible work associated with the nonadiabatic driving (and the corresponding increase in polarization) along these branches is given by,
\begin{equation}
W_{irr,i}=\int_0^{\tau_i}\omega\dot{P}dt=\frac{\omega_f+\omega_i}{2}\frac{\sigma^2}{\tau_i}.
\end{equation}
The total work done during a compression or expansion step is therefore  the sum,
\begin{equation}
\begin{aligned}
W_i^{tot}&=\left(\omega_f-\omega_i\right)\left[\frac{\sigma^2}{2\tau_i}+P_0\right]+\frac{\omega_f+\omega_i}{2}\frac{\sigma^2}{\tau_i}\\
&=\left(\omega_f-\omega_i\right)P_0+\frac{\sigma^2}{\tau_i}\omega_f.
\end{aligned}
\end{equation}
The work produced by the engine during one cycle, Eq.~(2),  is the sum over the compression and expansion steps  (with proper time matching) \cite{fel00}.

\subsection{Correlation functions of the two-level system}

The two-time correlation functions (9) of the two-level system may be obtained in the following way. We introduce the vector $\vec{A}$ having for components the three Pauli operators and the unit operator. The latter quantities form a complete set of observables for the two-level system. The master equation (1) can then be rewritten as a matrix differential equation, $d\vec{A}/dt=M\cdot \vec{A}$, or explicitly,
\begin{equation}
\frac{d}{dt}\begin{pmatrix}
 \sigma_x\\
\sigma_y\\
\sigma_z\\
I
\end{pmatrix}=
\begin{pmatrix}
-\frac{\gamma}{2}&-\omega&0&0\\
\omega&-\frac{\gamma}{2}&0&0\\
0&0&-\gamma&-\gamma_0\\
0&0&0&0
\end{pmatrix}
\begin{pmatrix}
\sigma_x\\
\sigma_y\\
\sigma_z\\
I
\end{pmatrix} .
\end{equation}
According to the quantum regression theorem \cite{bre07}, the equation of motion for the two-time correlation functions is the same as that for the operators, and we  thus have, 
\begin{equation}
\frac{d}{d\tau}\langle \hat{O}(t)\vec{A}(t+\tau)\rangle=M\langle \hat{O}(t)\vec{A}(t+\tau)\rangle.
\end{equation}
Since we are only interested in the correlation functions of the operator $\sigma_x$, we can restrict ourselves to the upper left $2\times 2$ submatrix of $M$. We find,
\begin{equation}
\label{eq}
\frac{d}{d\tau}\vec{C}=\begin{pmatrix}-\frac{\gamma}{2}&-\omega\\
\omega&-\frac{\gamma}{2}\end{pmatrix}\vec{C}(\tau),
\end{equation}
with the correlation vector, 
\begin{equation}
\vec{C}(\tau)=\begin{pmatrix}\langle \sigma_x(t)\sigma_x(t+\tau)\rangle\\\langle \sigma_x(t)\sigma_y(t+\tau)\rangle\end{pmatrix}.
\end{equation}
The solution to Eq.~\eqref{eq} is given by,
\begin{equation}
\vec{C}(\tau)=\left(
\begin{array}{cc}
 e^{-\frac{\gamma \tau}{2} } \cos ( \omega \tau ) & -e^{-\frac{\gamma t}{2} } \sin( \omega \tau ) \\
e^{-\frac{\gamma \tau}{2} } \sin ( \omega \tau )  & e^{-\frac{\gamma t}{2} } \cos ( \omega \tau )  \\
\end{array}
\right)\cdot \vec{C}(\tau=0),
\end{equation}
with the initial condition, 
\begin{equation}
\vec{C}(\tau=0)=\begin{pmatrix}\langle \sigma_x \sigma_x\rangle (t)\\ \langle \sigma_x \sigma_y\rangle (t)\end{pmatrix}=\begin{pmatrix} 1\\ i\langle \sigma_z\rangle (t)\end{pmatrix}.
\end{equation}
The last equality follows from the algebraic properties of the Pauli matrices. The symmetrized correlation function is equal to the real part of the above correlation function and reads,
\begin{equation}
C_{ij}=\frac{1}{2}\langle \left[\sigma_x(t+\tau),\sigma_x(t)\right]\rangle=e^{-\frac{\gamma}{2}\tau}\cos\left(\omega \tau\right).
\end{equation}


\begin{thebibliography}{99}
\bibitem{rem} Large nonadiabatic effects characterized by a small driving time $\tau_2$ are equivalent to a large $\bar \sigma= \sigma/\tau_2$.
\end{thebibliography}
\end{document}